\documentclass[twocolumn,twocolappendix,times]{aastex6}
\usepackage{amsmath}
\usepackage{bm}
\usepackage{times}
\newcommand{\ez}{{\bm{\hat{e}}_z}}

\defcitealias{2015ApJ...814...37M}{MH15}
\defcitealias{2013ApJ...775..114M}{MC13}

\begin{document}

\title{Photophoretic Levitation and Trapping of Dust in the Inner Regions of Protoplanetary Disks}
\shorttitle{Photophoretic Levitation of Dust in Disks}

\author{Colin P.~McNally\altaffilmark{1}}
\affil{Niels Bohr International Academy, The Niels Bohr Institute, University of Copenhagen, Blegdamsvej 17, 2100 Copenhagen \O, Denmark}
\author{Melissa K.~McClure}
\affil{European Southern Observatory, Karl-Schwarzschild-Str. 2, D-85748, Garching bei M\"{u}nchen, Germany}

\altaffiltext{1}{current address: Astronomy Unit, Queen Mary University of London, Mile End Rd, London, E1 4NS, UK}
\email{{\tt c.mcnally@qmul.ac.uk}, {\tt mmcclure@eso.org}}

\begin{abstract} 
 In protoplanetary disks, the differential gravity-driven settling of dust grains with respect to gas and with respect to grains of varying sizes
 determines the observability of grains, and sets the conditions for grain growth and eventually planet formation.
 In this work we explore the effect of photophoresis on the settling of large dust grains in 
 the inner regions of actively accreting protoplanetary disks.
 Photophoretic forces on dust grains result from the collision of gas molecules with differentially heated  grains.
 We undertake one-dimensional dust settling calculations to determine the equilibrium vertical distribution of dust grains in each column of the disk.
 In the process we introduce a new treatment of the photophoresis force which is consistent at all optical depths with the 
 representation of the radiative intensity field in a two-stream radiative transfer approximation.
 The levitation of large dust grains creates a photophoretic dust trap several scale heights above the mid-plane
in the inner regions of the disk where the dissipation of accretion energy  is significant. 
We find that differential settling of dust grains is radically altered in these regions of the disk, 
with large dust grains trapped in a layer below the stellar irradiation surface, in where the 
dust to gas mass ratio can be enhanced by a factor of a hundred for the relevant particles.
The photophoretic trapping effect has a strong dependence on particle size and porosity.

\end{abstract}


\section{Introduction}

Dust in circumstellar or protoplanetary disks around T-Tauri stars contributes to an 
observed infrared excess in the spectral energy distribution of these 
sources and provides the formation material for planetary systems.
The vertical distribution of dust has observational consequences and is critical for planet formation
consequences in these disks \citep{2014prpl.conf..339T}.
There are several processes at play in the inner disk that
can simultaneously affect the dust distribution at a given radius.
\citet{1977MNRAS.180...57W,1980Icar...44..172W} considered the 
competition between gravity and turbulence in setting the thickness 
of the dust sub-disk in the solar nebula as a precursor for planet formation.
The beginning of what could be termed the modern treatment of the equilibrium between 
vertical turbulent diffusion and gravity for the dust distribution, also known as the dust settling,
was in the analytical solutions for the equilibrium dust distribution given by \citet{1995Icar..114..237D}.
The work of \citet{2004A&A...421.1075D} solved for the time evolution
of the dust distribution, to determine if specific grain sizes can be 
expected to settle during the lifetime of a observed disk.
This previous work has thus been concerned with characterizing the 
appropriate form for the vertical motion of dust,
and finding the equilibrium and the time evolution.

A third process, that can affect the distribution of dust is photophoresis. 
The form of photophoresis \citep{1918AnP...361...81E} 
that we are concerned with in this work is the ``$\Delta T_s$-photophoresis'',
where temperature differences established across a grain by  
directional differences in incident light radiation, result in 
a force due to the interaction with the gas.
Gas molecules reflecting off the hotter side of the particle pick up more 
energy and momentum from the hot side of the particle than the cold side,
resulting in a net force on the particle directed away from the brighter incident radiation
\citep{1967JGR....72..455H,rohatscheck1985,mackowski1989,1993PhFl....5.2043B,rohatscheck1995}.
The reader is reminded that this force is not radiation pressure, 
which results from the momentum of interacting photons.

The role of photophoresis driven by direct stellar radiation in 
optically thin regions of the disk has been studied in a number of contexts
\citep{2005ApJ...630.1088K,2006Icar..180..487W,2007A&A...462..977K,2008ApJ...677.1309T,2016Icar..268..281C,2016MNRAS.458.2140C}.
Photophoretic levitation of dust above the disk 
surface during FU~Orionis outbursts has been examined by 
\citet{2009M&PS...44..689W}, \citet{2009ASPC..414..509W}, in a different form and regime. This 
is an optically thin case where the dust is forced high 
enough that the particle is exposed to direct optically thin
 irradiation from the star and disk while the disk is in an outburst state.
 The resulting focus of that work is on the radial
  transport of large calcium-aluminum-rich inclusions in a
  layer termed an equilibrium corridor.
 Importantly, \citet{2009M&PS...44..689W} recognized that the maximum height 
 to which particles can be photophoretically levitated above the disk to is set primarily 
 by the decreasing gas density with height, that rapidly attenuates the vertical force.

In contrast, here we follow up the suggestion in \citet{2015ApJ...814...37M}, 
referred to as \citetalias{2015ApJ...814...37M}, 
 that dust can be levitated by 
photophoresis below the stellar irradiation
 surface of the disk at moderate and high optical depths,  
 driven solely by the disk's own diffuse thermal radiation.
 We extend the treatment of photophoresis from \citetalias{2015ApJ...814...37M} 
to one consistent with the two-stream radiative transfer approximation used in \citet[][referred to as \citetalias{2013ApJ...775..114M}]{2013ApJ...775..114M}.
Using disk models matched to observations from \citetalias{2013ApJ...775..114M},
that include radiative transfer, vertical hydrostatic equilibrium, and viscous heating, we calculate the photophoretic force on the particles.
Then,  we solve for the evolution and equilibrium of the vertical dust distribution 
in a protoplanetary disk under the combined influence of gravity, gas turbulence, and photophoresis, extending the method of \citet{2004A&A...421.1075D}
to include the latter force.

In Section~\ref{sec_methods} we present the methods used in our calculations.
Section~\ref{sec_twostream} is devoted to introducing the novel formulation of 
photophoresis in the context of a two-stream approximation of radiative transfer used in this work.
In Section~\ref{sec_citau} we give results for a range of calculations preformed in the context of CI~Tau,
in Section~\ref{sec_v836} we briefly comment on V836~Tau,
and finally Sections~\ref{sec_disc} and \ref{sec_conc} contain discussion of the results and conclusions, respectively.

\section{Methods}
\label{sec_methods}
The underlying disk models are calculated with an updated version of the \citet{2006ApJ...638..314D} code, derived from \citet{1998ApJ...500..411D}. 
The models used are those for CI~Tau and V836~Tau from \citetalias{2013ApJ...775..114M}.
From these models, we have extracted the grid of gas density ($\rho_g$) along with the complete description of the 
approximate radiation field: the mean intensity ($J_{\rm rad}$) and net vertical flux of radiation ($F_{\rm rad}$).

With this as the background, describing the gas and radiation in the disk, we perform a dust settling calculation in the style of 
\citet{2004A&A...421.1075D} in vertical columns in the disk.
The conservation equation for the dust particle number volume density $n$ is 
\begin{align}
\frac{\partial n}{\partial z} - \frac{\partial}{\partial z}\left[\rho_g D \frac{\partial }{\partial z}\left(\frac{n}{\rho_g}\right)\right]
 +\frac{\partial}{\partial z}\left(n v_d\right) = 0
 \label{eq_dustevo}
\end{align}
where $z$ is the distance above the disk midplane, $D$ is the dust diffusion coefficient, and $v_d$ is the dust drift velocity \citep{1995Icar..114..237D}.
Although this equation specifies the time evolution of the dust distribution, 
in this work we restrict our attention to steady-state solutions.
The equilibrium dust distribution is calculated following \citet{2004A&A...421.1075D}
with the modification that the advective 
derivative is upwinded, depending on the direction of the net force.

Dust diffusion is a consequence of the turbulent (viscous) angular momentum transport and energy
dissipation in the underlying disk model. 
The reader should note that the literature on determining the relations, constant, and scalings 
has become complicated due to the use of a wide array of different definitions and assumptions.
In terms of the turbulent $\alpha_{\rm ss}$  \citep{1973A&A....24..337S} the radial diffusion of angular momentum $\nu$ is
\begin{align}
\nu = \alpha_{\rm ss}\frac{ c_s^2 }{ \Omega_K}
\end{align}
where $c_s$ is the sound speed and $\Omega_K= \sqrt{GM_\star/R^3}$ is the
 Keplerian orbital frequency around a star with mass $M_\star$ at cylindrical radius $R$.
 Note that the \citetalias{2013ApJ...775..114M} models assume a single constant value of  $\alpha_{\rm ss}$ for each disk.
We follow the convention often used in numerical experiments that the dust vertical diffusivity is
\begin{align}
D = \frac{\nu}{\rm Sc} \label{eq_sc}
\end{align}
which serves to define ${\rm Sc}$ as the vertical Schmidt number.\footnote{Many similar but distinct definitions and terminology exist,
 in particular similar quantities have been  referred to as an
(effective) Prandtl number \citep[ex:][]{2005MNRAS.358.1055C}.}
Here, the vertical Schmidt number parameterizes the relative radial diffusion of angular 
momentum to the vertical diffusion of dust particles.
The reader should be aware that in other works, the vertical Schmidt number is also defined 
as the ratio of vertical gas particle diffusion to vertical dust particle diffusion \citep{1993Icar..106..102C,2007Icar..192..588Y}.
Despite the variations in the definition, it is expected to be close to unity for dust particles with small stopping times.
However, as the results in this work have a very strong dependence on $\rm Sc$, 
in particular with small variations from unity, the precise definition is critical.
This arises because the vertical Schmidt number connects the vertical diffusion of the dust $D$ 
to the accretion luminosity of the disk given through $\nu$ and thereby to any photophoretic levitating force.

The stopping time is captured by ${\rm St}$, the Stokes number of the dust particle, being defined as
\begin{align}
{\rm St} = \Omega t_s
\end{align}
(in the terms of \citet{2007Icar..192..588Y} this assumes the eddy timescale $t_e=\Omega^{-1}$).
In this paper the particles treated are overwhlemingly in the $\rm St \rightarrow 0$ limit by 
the time they have settled to their equilibrium distribution. Only when particles are initially 
falling from the well-mixed initial condition used in the calculations do some have $\rm St \sim1$ at large altitudes.

Simulations of magnetorotational instability driven 
 turbulent disk flows suggest that, for particles in the vanishing Stokes number 
limit, $\rm Sc$ is a factor of a few larger than unity and can differ 
significantly from the horizontal Schmidt number 
\citep{2005MNRAS.358.1055C,2005ApJ...634.1353J,2006ApJ...639.1218T,2009A&A...496..597F,2010MNRAS.409..639N,2015ApJ...801...81Z}. 
Moreover, as the stopping time grows to be significant, the Stokes number grows, and 
the vertical Schmidt number 
again varies.
To capture both of these we combine a limiting value with 
 the dependence on the Stokes number given by \citet{2007Icar..192..588Y} as
\begin{align}
{\rm Sc} = {\rm Sc_{0}}(1+{\rm St}^2)
\end{align}
where ${\rm Sc_{0}}$ is the limiting value as ${\rm St}\rightarrow 0$.
The constant ${\rm Sc_{0}}$ is uncertain, but our results will have a important dependence on it.
For a discussion of the physical origin of a Schmidt number different from unity in magnetorotational 
instability driven accretion disk turbulence the reader is referred to \citet{2009A&A...496..597F}.
In the calculations in this paper, the Stokes numbers for particles affected by photophoresis are very small, as
photophoresis arises from collisions with the gas molecules, 
so it operates best in the vanishing Stokes number limit.

The stopping time $t_s$ is determined by the drag force on the dust particle.
In this work the dust particles have sizes much less than the gas molecule mean free path
and drift velocities much less than the gas molecule thermal velocity, so 
we use the expression for the drag force $\bm{F}_{\rm drag}$ in the Epstein regime 
for spherical particles
given by \citet{2003A&A...399..297W} as
\begin{align}
\bm{F}_{\rm drag} = -\frac{8\sqrt{\pi}}{3} a^2 \rho_g  v_{\rm T} \bm{v}
\end{align}
where $a$ is the particle radius, $v_{\rm T}=\sqrt{2 k_B T_g / \mu m_H} $ is the thermal velocity of gas molecules, 
$\bm{v}$ is the dust particle velocity,
$k_B$ is the Boltzmann constant, $T_g$ is the gas temperature, $m_H$ is the mass of one $\rm amu$, 
and $\mu$ is the average gas molecule mass in $\rm amu$. 
The stopping time for a dust particle with mass $m_d$ is then defined by
\begin{align}
\frac{d {\bm v}}{dt} = - \frac{\bm{v}}{t_s} = \frac{{\bm F}_{\rm drag}}{m_d}\ .
\end{align}

The final term in Equation~(\ref{eq_dustevo}) advects the dust vertically at the drift velocity $v_d$.
This is defined as the terminal drift speed where the vertical gravity, photophoresis, and drag forces balance
\begin{align}
F_{g} + F_{p} = F_{\rm drag} \ .
\end{align}
The terminal velocity approximation is appropriate where the stopping time $t_s$ is small 
compared to the particle's free fall time.
The vertical component of the central star's gravity $F_{\rm g}$ on the dust particle is
\begin{align}
F_{\rm g} = -\frac{G M_\star m_d z}{(R^2+z^2)^{3/2}}\ .
\end{align}
where $M_\star$ is the stellar mass.
The appropriate form for the photophoretic force $F_{\rm ph}$ is presented in the next section.

\subsection{Consistency of the Modeling Approach}
\label{sec_preconsistency}
The approach used in this work is to use the thermal and radiative transfer structure 
from the \citetalias{2013ApJ...775..114M} models and post-process dust settling calculations overtop of this.
The \citetalias{2013ApJ...775..114M} models themselves already contain a prescription for the 
vertical distribution of small and large dust,
so the combined models in this work are only strictly self-consistent if the dust population which
is distributed differently in the separate settling calculations in this work make 
a small contribution to the total Rosseland mean opacity.
Additionally, in this work, the calculations only present the relative distribution of 
dust of a single particle type in a  given column of the disk, and the absolute 
quantity and overall composition of the dust in the disk is not specified.
Discussion of such further extensions to the approach, in light of the results 
obtained in this work, are postponed until Section~\ref{sec_further}.

\section{Photophoresis with Two-Stream Radiative Transfer}
\label{sec_twostream}

Although the midplane of the inner parts of a protoplanetary disk
are optically thick, the optical depth decreases with height. 
If particles are photophoretically lofted high enough, they will be moved beyond the region where the 
optically thick approximation of \citetalias{2015ApJ...814...37M} strictly applies. 
The optically thick approximation in \citetalias{2015ApJ...814...37M} matches this two-stream approach 
at high optical depths near the midplane, but underestimates and eventually gives
 the wrong sign for the photophoretic force in the vertical direction at high altitudes, 
 where the vertical gradient of gas temperature reverses, while the direction of the radiative flux remains the same.
 As we find that dust particles do indeed reach such altitudes, we develop here the required expressions 
using the results of a two-stream radiative  transfer calculation to derive the photophoretic force.
Since the underlying disk models used in this work use the two-stream approximation on vertical rays for the radiation field, the
photophoretic force derived in this way is exactly consistent with the radiation field in those models.
Again, the medium is considered to be dilute, so that the gas molecule mean free path is
much larger than the size of the particles experiencing the photophoretic force.
Gas molecule mean free paths can be estimated as \citep{2009AnRFM..41..283S}
\begin{align}
\lambda_{\rm mfp} = 1 \ \left(\frac{\rho_g}{10^{-9}\ \mathrm{g\ cm^{-3}}}\right) \ \mathrm{cm}
\end{align}
which, in the regions of the disks under consideration is 
comfortably larger than the particles' sizes considered.
The opacity of the medium is considered to be dominated by small dust grains, 
of size comparable to and smaller than the wavelengths of the  radiation field.
The analysis of the photophoretic force here is again appropriate for opaque particles 
with low reflectivity at the appropriate wavelengths much 
larger than the wavelengths of the radiation field.

In the two-stream radiative transfer approximation applied to a plane parallel atmosphere,
 the radiation intensity field is approximated
by its first and second moments, or equivalently the mean intensity 
$J$ and net flux $F$ in the $z$ direction as
\begin{align}
I(z,\theta) = J(z) - \frac{3}{4\pi}F(z) \cos(\theta) \label{eq_2streami} \ ,
\end{align}
with $\theta$ being the angle away form the positive $z$ axis.
Note, this expression in the optically thick limit reduces to the expression given in \citetalias{2015ApJ...814...37M} by their Equation~(10).
The linearization and solution of the problem proceed in the same way as in
\citetalias{2015ApJ...814...37M} their Section~2.2 and~2.3, so we do not repeat those steps here.
The constant terms yield the relation
\begin{align}
\sigma_{\rm SB} T_0^4 +\Upsilon T_g^{1/2} T_0  -\Upsilon T_g^{1/2}T_g -\pi J = 0
\end{align}
where $\Upsilon$ is the coefficient for heat conduction to the gas given by \citetalias{2015ApJ...814...37M} in their Equation~(17). This expression can be solved analytically for $T_0$.
However, the analysis of the photophoretic force is made in the limit that $T_0\approx T_g$,
so we will proceed using that approximation, as it is generally accurate at moderate and 
high optical depths.\footnote{Because of this, the additional effects considered by \citet{2016arXiv160501022L} are sub-dominant here.}
The $\cos(\theta)$ angular dependence yields the relation
\begin{align}
A_1 &= -\frac{1}{2} F \left( k + 4\sigma_{\rm SB} T_0^3 a + \Upsilon T_g^{1/2} a \right)^{-1}
\end{align}
where $k$ is the thermal conductivity of the particle.
Applying the approximation $T_0\approx T_g$ and using Equation~(7) of \citetalias{2015ApJ...814...37M} gives the photophoretic force as
\begin{align}
\bm{F}_p &\approx \frac{\pi}{6} \alpha \frac{k_B \rho_g}{\mu m_H} \frac{a^3}{k} F \left( 1 + 4\sigma_{\rm SB} T_g^3 \frac{a}{k} + \Upsilon T_g^{1/2} \frac{a}{k} \right)^{-1} \ez \label{eq_2streamfp}
\end{align}
where $\alpha$ is the surface accommodation coefficient which we take to be $1$ as in \citetalias{2015ApJ...814...37M}.
This expression can be compared to Equation~(26) of \citetalias{2015ApJ...814...37M}, which gives the optically thick limit.
Note that in Equation~(\ref{eq_2streamfp}) $\ez$ is strictly defined as the direction of the radiation streams in the two-stream approximation.
The limitations of the accuracy of the photophoretic force, due to the linear approximation of the force integral 
over the particle surface, are similar.

As an approximation, Equation~(\ref{eq_2streamfp}) is best in the optically thick 
and moderate optical depth regimes where Equation~(\ref{eq_2streami}) is a good 
representation of the radiative intensity field.
Although the two-stream approximation produces the correct net radiative flux $F$ at all optical depths,
Equation~(\ref{eq_2streamfp}) does not capture the possible one-sided nature of 
direct illumination which can occur at very low optical depths. We computed
numerical solutions of the heat transfer problem, for a sphere illuminated from a single direction
at low optical depth, and found  that in that worst case Equation~(\ref{eq_2streamfp}) 
underestimates the true force by approximately a factor of two.
However, in the application presented in this paper such a situation does not occur.

The thermal conductivity of the particles plays an important role in the strength of the 
photophoretic force (\citealt{2012A&A...545A..36L}, \citetalias{2015ApJ...814...37M}, \citealt{2016MNRAS.458.2140C})
Here, we use the same formulation as in \citetalias{2015ApJ...814...37M}.
This is based on the experiments for packed silicate aggregates of \citet{2011Icar..214..286K} and has the form
$k=k_0\exp(7.91\phi)$ where $k_0=51.4\ \mathrm{erg\ s^{-1}\ cm^{-1}\ K^{-1}}$, and
$\phi$ is the grain's volume filling factor. The volume filling factor is defined for a given sample as
$\phi\equiv \rho_d/\rho_0$ where $\rho_d$ is the density of the bulk sample, and $\rho_0$ is the 
mass density of the constituent solid silicate monomers. 

The spin of dust particles can damp the photophoresis effect if it is fast enough so that a
thermal gradient fails to develop across the particle. For the dust particles studied 
in this work, the rotation period of dust is expected to be slow compared to the thermal 
conduction timescale, so photophoresis is efficient. Details of this analysis are given in the appendix.

\section{Results for CI Tau}
\label{sec_citau}

\begin{table}
\caption{Parameters of CI Tau Model}
\label{tab_params}
\begin{center}
\begin{tabular}{lc}
\hline
\hline
Parameter & Value\\
\hline
Stellar Mass (${ M_\sun}$) & $0.8$ \\
Accretion Rate (${ M_\sun\ yr^{-1}}$) &  $2.9\times10^{-8}$\\
Disk Mass (${ M_\sun}$) & $6.8\times 10^{-2}$ \\
Inner Disk Wall radius, midplane ($\mathrm{au}$) & 0.12\\
$\alpha_{\rm SS}$ & $5\times10^{-3}$\\
\hline
\end{tabular}
\end{center}
\end{table}

\begin{figure}
\includegraphics[width=\columnwidth]{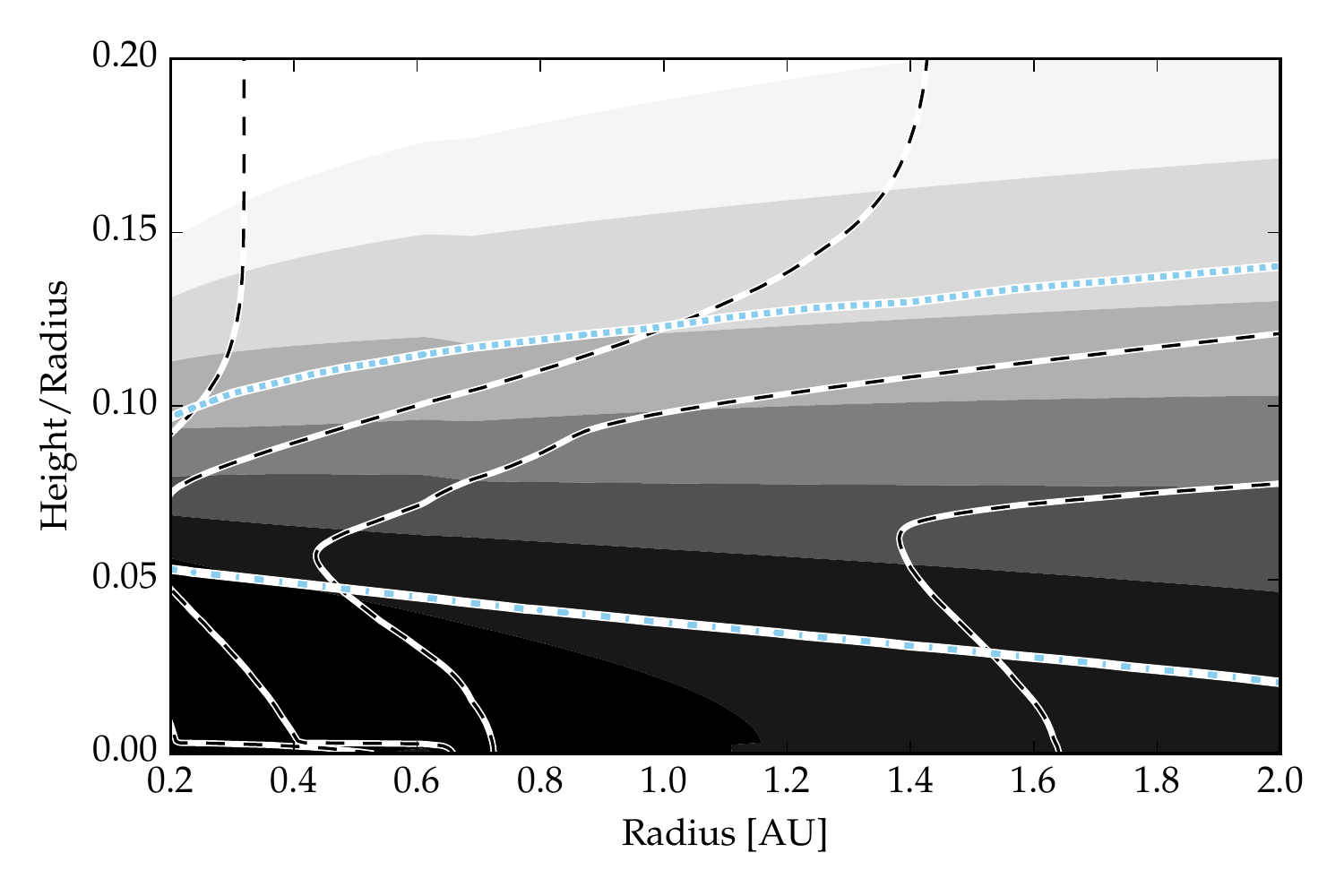}
\caption{Section of inner disk structure of CI Tau model from \citet{2013ApJ...775..114M}. 
Filled contours: $\rho_g$, from top to bottom $10^{-15}$, $10^{-14}$, $10^{-13}$, $10^{-12}$, $10^{-11}$, $10^{-10}$, $10^{-9}$ $\rm g\ cm^{-3}$. 
Dashed contours: $T_g$, from left to right $800$,  $400$, $200$, $100\ {\rm K}$.
Dotted light blue contour: $z_s$ surface where the optical depth to incoming stellar irradiation is $1$.
Dot-dash light blue contour: $z_{\rm phot}$  disk photosphere surface where the optical depth to outgoing disk thermal radiation is $2/3$.}
\label{fig_citau_disk_struct}
\end{figure}

CI Tau is a `typical' classical T Tauri system undergoing an average amount of accretion from the disk onto the star. Vital parameters 
from the best fit model of \citetalias{2013ApJ...775..114M} as used in this 
analysis, are reproduced in Table~\ref{tab_params}.
Figure~\ref{fig_citau_disk_struct} gives the density and temperature
structure of the region of the disk  model from \citetalias{2013ApJ...775..114M} studied here.
Along with the density and temperature structures, two surfaces relating to the radiation are shown.
The disk photosphere, as defined by the height $z_{\rm phot}$ where the optical depth to
 outgoing disk thermal radiation is $2/3$ 
in the Rosseland mean, and the surface where incoming stellar irradiation is scattered or absorbed at $z_s$, 
where the optical depth to incoming stellar irradiation is $1$ in the Planck mean \citep{1998ApJ...500..411D}.
The disk is heated both by viscous heating and stellar irradiation. 
In this radial range, the midplane is dominantly heated by viscous dissipation, and the 
temperature decreases through the photosphere, 
from where thermal radiation can freely escape.
Above this, the disk atmosphere temperature rises again as the combined 
effects of the low opacity of the gas to thermal radiation prevents it from cooling 
efficiently, and the stellar irradiation, which penetrates down to $z_s$, heats the atmosphere.
The relative vertical and radial gradients of the radiation field can be 
crudely judged by the slope of the disk photosphere, 
which deviates less than one degree from parallel to the midplane in the region shown. 
Hence, the radiation field in this region can indeed be well approximated as 
dominantly vertical, perpendicular to the midplane.
Two dust populations are included in this model; 
a population of small dust which is well-mixed with the gas,  and a distribution 
extending to large sizes which is settled to a thin layer with a dust scale 
height one tenth of the gas pressure scale height. 
The mass of settled dust is assumed to be vertically conserved, so the dust/gas ratio 
is decreased in the upper layers and proportionally increased in the midplane \citep{2006ApJ...638..314D}.
Due to its inclusion of viscous heating by accretion, 
 parametrized by an $\alpha$-viscosity $\alpha_{\rm SS}$, the disk
 temperature structure is dominated by a vertical temperature inversion at small radii. 
 The gas temperature at moderate disk altitudes is lower than either the hot midplane below, 
 heated by viscous dissipation, or the warm atmosphere slightly above, heated by stellar irradiation.
As heat in the form of radiation can only escape vertically, the net flux of thermal 
radiation must be directed vertically up everywhere, resulting in a photophoresis force 
directed up, away from the midplane.

\subsection{A Photophoretic Dust Trap}
\begin{figure}
\includegraphics[width=\columnwidth]{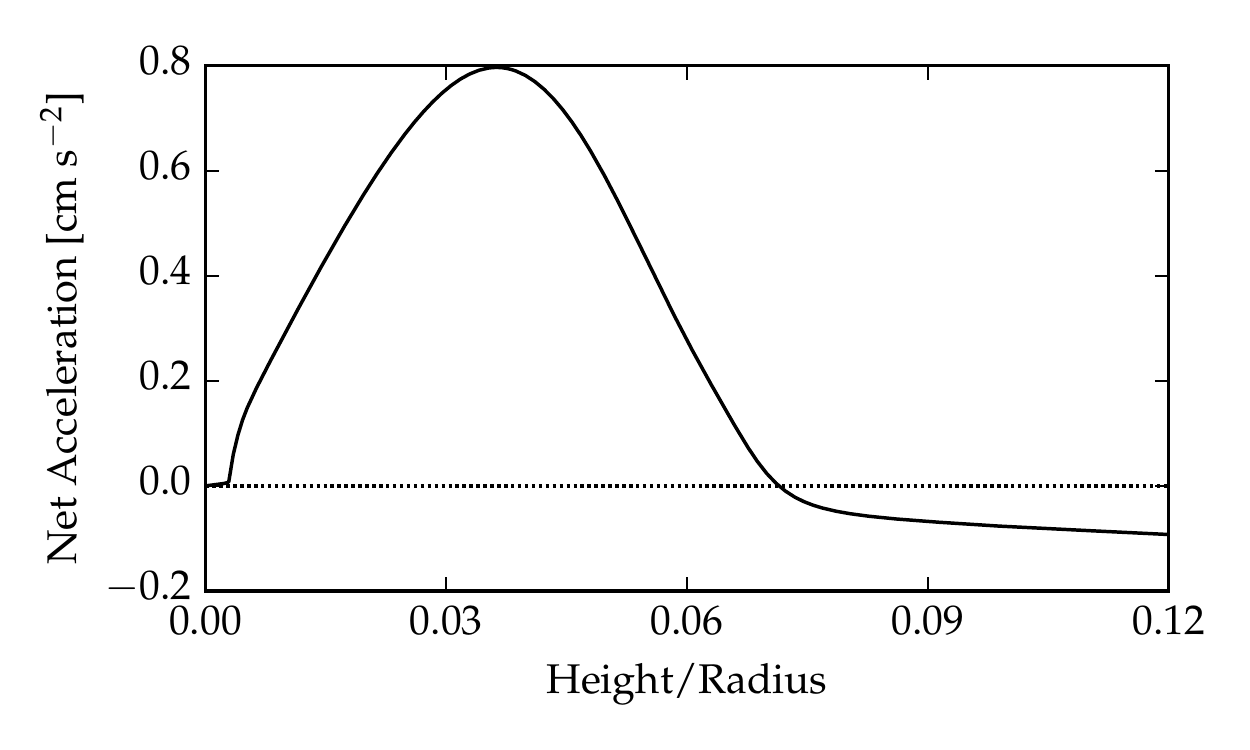}  
\caption{Net vertical acceleration (gravity and photophoresis) for dust particles with $a=1\times10^{-2}\ \mathrm{cm}$, $\phi=1$, at $R=0.78\ \mathrm{au}$.
Below Height/Radius $\sim 0.07$ photophoresis dominates over vertical gravity forcing particles upwards, 
until the decrease in gas density in the disk atmosphere attenuates the force, forming a trap.}
\label{fig_force_a1e-2_phi1_sc1p5}
\end{figure}

\begin{figure*}
\includegraphics[width=\textwidth]{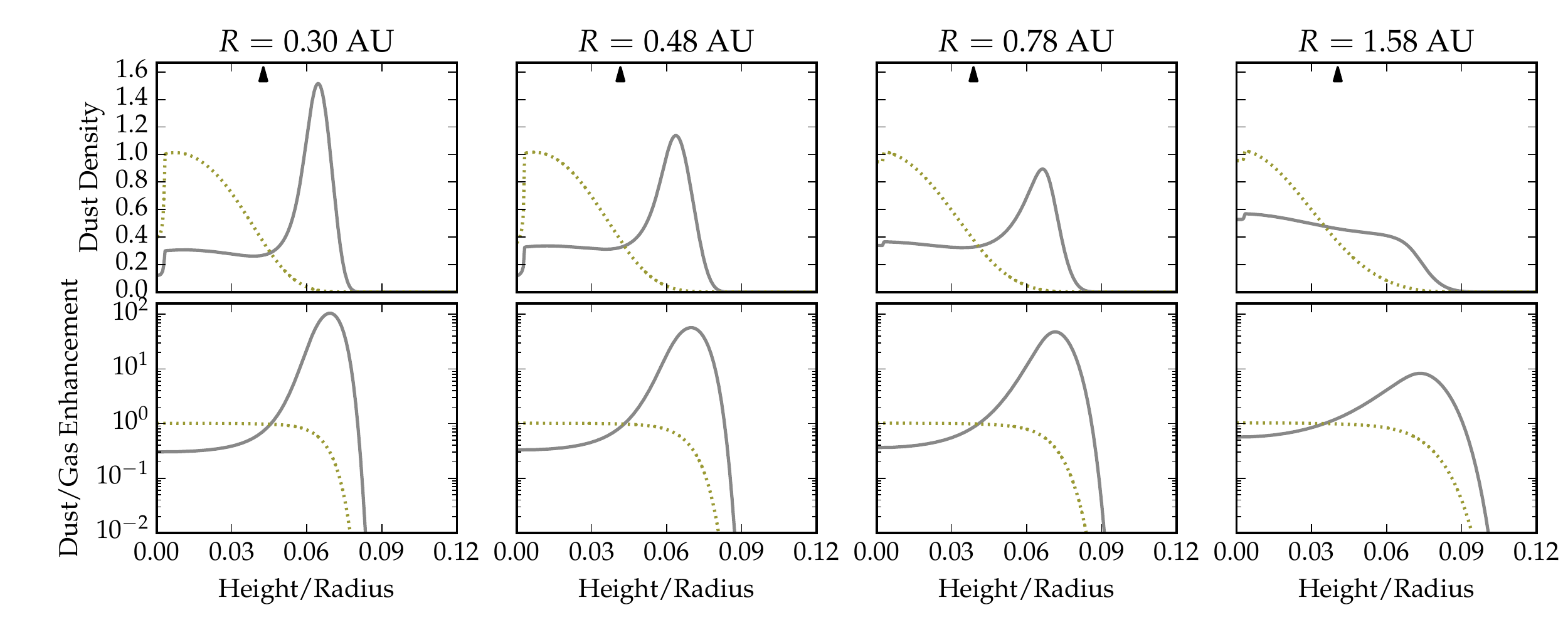}  
\caption{Fiducial case: equilibrium vertical distribution of dust particles with $a=1\times10^{-2}\ \mathrm{cm}$, $\phi=1$, with ${\rm Sc_0}=1.5$. 
Top row: density relative to the well-mixed peak density on a linear scale, the black triangle marks one gas density scale height.
Bottom row: dust to gas density ratio enhancement factor over the well-mixed density on a logarithmic scale. 
Green dotted line: turbulence and gravity only. Gray solid line: including photophoresis, this is the fiducial case repeated in later figures.}
\label{fig_a1e-2_phi1_sc1p5}
\end{figure*}

The upwards photophoretic force grows from zero at the midplane as 
the vertical flux of radiation increases
and is rapidly cut off with increasing altitude, not by the 
properties of the disk radiation field, but by the rapidly decreasing gas density
in the disk atmosphere. Thus, while the details of the radiation field are critical near the midplane, 
the gas density dominates above the disk photosphere.
At the gas density steeply attenuates the upwards photophoretic force, the vertical 
component of the central star's gravity is increasing linearly with altitude.
This situation is demonstrated in Figure~\ref{fig_force_a1e-2_phi1_sc1p5},
where the net acceleration of an example dust particle is shown. 
It is positive from the 
midplane up to a stable equilibrium point at approximately Height/Radius$\ \sim0.07$,
and negative at larger altitudes where vertical gravity dominates.
Thus, a photophoretic trap for dust grains which experience photophoresis is formed at Height/Radius$\ \sim0.07$.

The drop off in the gas density in also critical to setting the dust settling height when 
considering grains which do not, or only weakly, experience photophoresis.
However, for the very small particles
the Stokes number at each height in the column is much lower than it is 
 for the particles large enough to experience photophoresis.
Hence, the large grains levitated by photophoresis are not expected to rise above the micron-sized 
gains which provide opacity. Large grains will be shielded from direct radiation from the central star by the small grains.
This situation is in strong contrast to that envisioned by  \citet{2009M&PS...44..689W} 
during FU~Orionis outbursts, where large grains are levitated above the opacity provided by
 small particles and remain directly exposed to the stellar irradiation.
Levitated large grains will still be in an environment where they
 collide with small grains, albeit at a lower frequency than at the midplane due to the lower density.
Similarly, the photophoretic trap for large grains forms below the surface where 
the optical depth to stellar irradiation is significant ($\tau_s=1$). Since the large grains are not directly exposed to the 
star's light, they are thus not expected to have a direct effect on the emission of scattered light from the disk.

Figure~\ref{fig_a1e-2_phi1_sc1p5} shows what we will take in this paper as the fiducial result
for the vertical settling problem given by Equation~\ref{eq_dustevo},
with solid ($\phi=1$) particles of radius $a=5\times10^{-3}\ \mathrm{cm}$ and the assumption ${\rm Sc_0}=1.5$.
Equilibrium solutions are shown, both with and without the photophoretic force.
We plot both the dust density normalized to the global peak density of a well-mixed 
(constant dust/gas mass ratio) dust, and the 
enhancement in the local dust/gas mass ratio over well-mixed dust.
The first gives the distribution of the dust in absolute terms on a linear $y$-axis scale, and the second gives
the local enhancement in the dust density over an hypothetical initial well-mixed state on a logarithmic $y$-axis scale, which is 
useful as a figure of merit for considering stability.
The result without photophoresis shows the conventional behavior of dust settling for particles with small Stokes number;
that particles settle out of the very lowest density regions, and sediment below, and moreover for these small grains, the 
dust remains to a good approximation well-mixed up to where it has settled out.

We find for this case the local dust/gas enhancement reaches fifty to one hundred for a wide range in the inner disk.
If the background well-mixed dust/gas mass ratio of dust grains which behave in this way (i.e.\ excluding small grains)
is $1\%$ then the local dust/gas mass ratio would be unity in the dust trap.
At levels even approaching this figure, instabilities of some manner can be expected \citep{2005ApJ...620..459Y,2006ApJ...643.1219J}.

\subsection{The Vertical Schmidt Number and $\rm Sc_0$}

\begin{figure*}
\includegraphics[width=\textwidth]{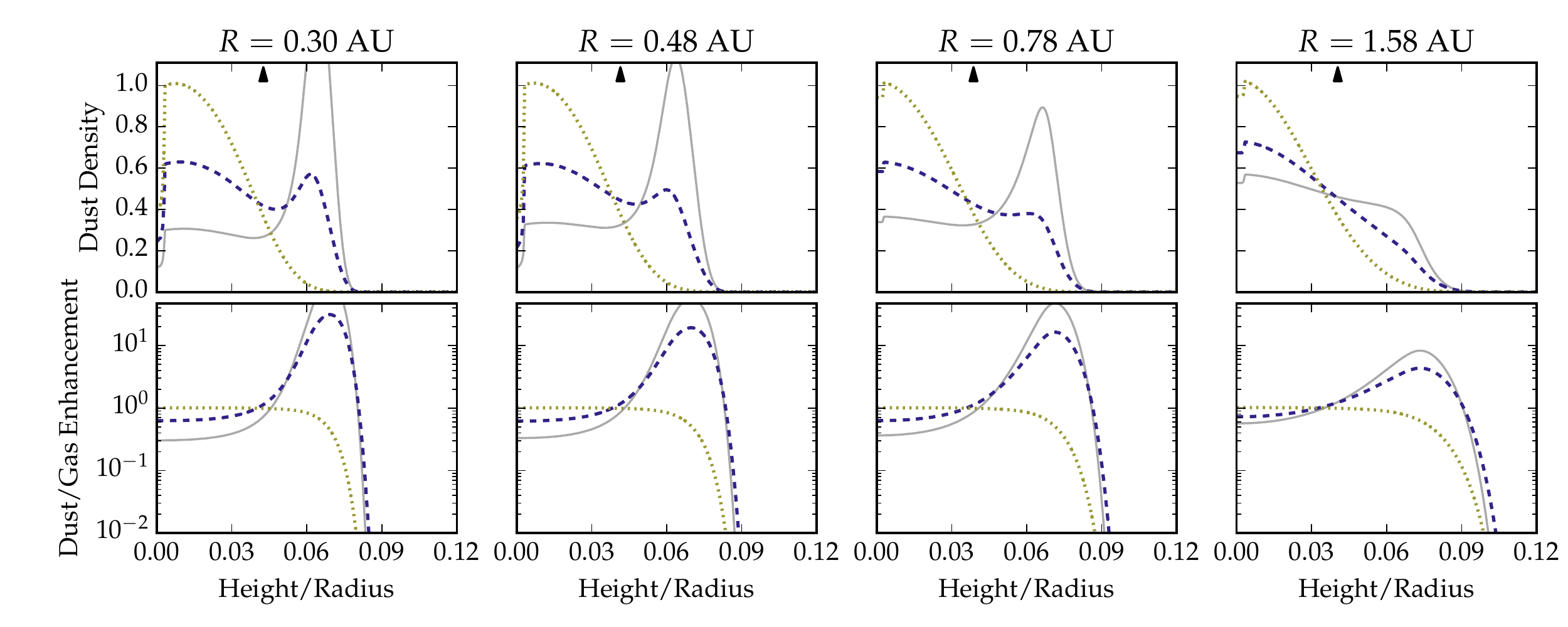}  
\caption{Schmidt number variation, ${\rm Sc_0}=1.0$ case:  equilibrium vertical distribution of dust particles with $a=1\times10^{-2}\ \mathrm{cm}$, $\phi=1$, with ${\rm Sc_0}=1.0$. Top row: density relative to the well-mixed peak density, the black triangle marks one gas density scale height.
Bottom row: dust to gas density ratio enhancement factor over the well-mixed density. Green dotted line: turbulence and gravity only. Blue dashed line: including photophoresis. Gray solid line: fiducial case.}
\label{fig_a1e-2_phi1_sc1p0}
\end{figure*}

\begin{figure*}
\includegraphics[width=\textwidth]{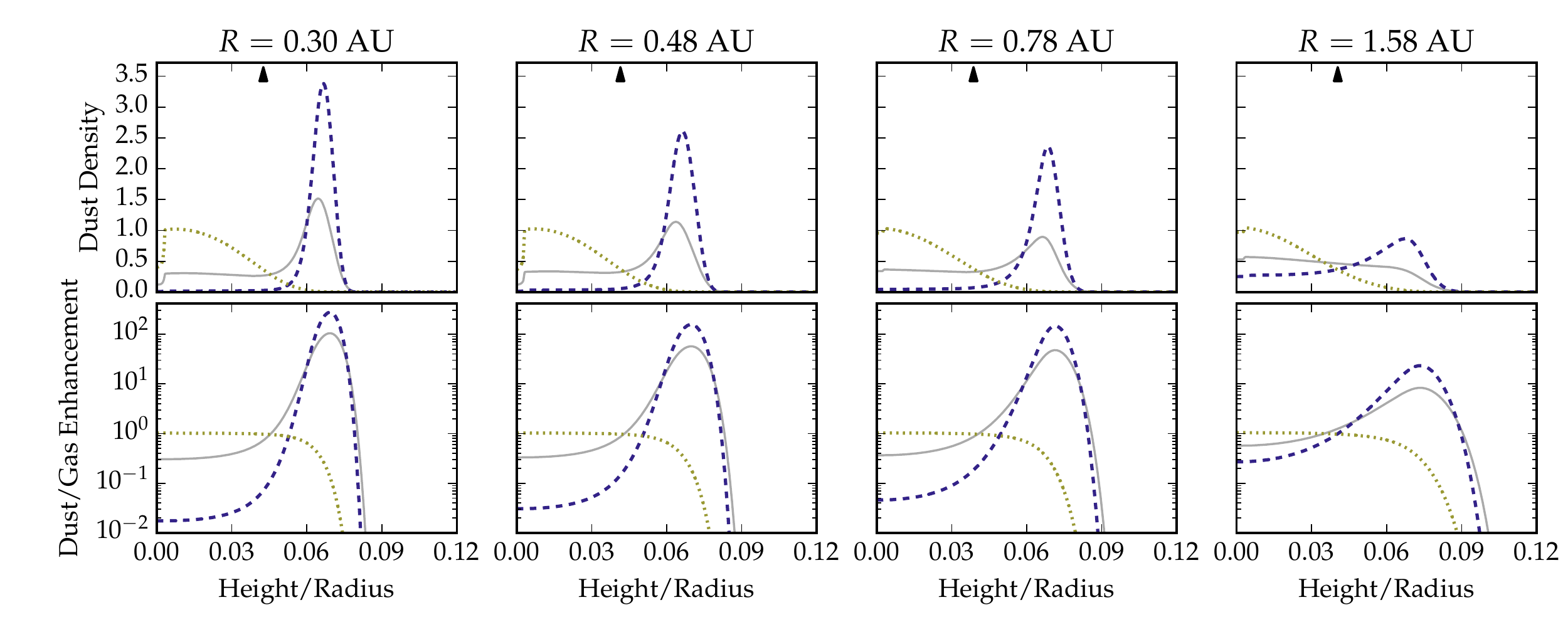}  
\caption{Schmidt number variation, ${\rm Sc_0}=2.5$ case: equilibrium vertical distribution of dust particles with $a=1\times10^{-2}\ \mathrm{cm}$, $\phi=1$. Top row: density relative to the well-mixed peak density, the black triangle marks one gas density scale height.
Bottom row: dust to gas density ratio enhancement factor over the well-mixed density. Green dotted line: turbulence and gravity only. Blue dashed line: including photophoresis. Gray solid line: fiducial case.}
\label{fig_a1e-2_phi1_sc2p5}
\end{figure*}

The vertical Schmidt number used in this work was introduced in 
Equation~(\ref{eq_sc}).
Numerical experiments attempting to determine the correct value for magnetorotational instability 
driven turbulence have a wide range of results 
\citep{2005MNRAS.358.1055C,2005ApJ...634.1353J,2006ApJ...639.1218T,2009A&A...496..597F,2010MNRAS.409..639N,2015ApJ...801...81Z}
and, as an additional complication, the unspecified instability leading to the $\alpha$-viscosity 
in the  \citetalias{2013ApJ...775..114M} models need not be magnetorotational instability.
Thus, we consider here a variation of $\rm Sc_0$ around the fiducial case of $1.5$ used in the previous section.
Note here, how we follow a definition in relation to the radial angular momentum transport.
The same calculation as in the previous section, but with ${\rm Sc_0} = 1.0$ is shown in Figure~\ref{fig_a1e-2_phi1_sc1p0} 
and that with ${\rm Sc_0} = 2.5$ is shown in Figure~\ref{fig_a1e-2_phi1_sc2p5}.
It is clear that the value of the Schmidt number adopted in this model has a significant impact on the trapping of dust.
Generally, a Schmidt number larger than unity decreases the tendency of turbulence to return the dust to a well-mixed state at the midplane, leaving more particles in the photophoretic trap and a lower dust density at the midplane,
as reflected dramatically in Figure~\ref{fig_a1e-2_phi1_sc2p5}.
We note that this strong dependence of the dust distribution on the vertical Schmidt number is quite 
different from the behavior in the case neglecting photophoresis, where the dust distribution 
barely varies as the vertical Schmidt number in changed in this range 
(see Figure~\ref{fig_a1e-2_phi1_sc1p0} and Figure~\ref{fig_a1e-2_phi1_sc2p5}, green dotted lines).

\subsection{Particle Size}

\begin{figure*}
\includegraphics[width=\textwidth]{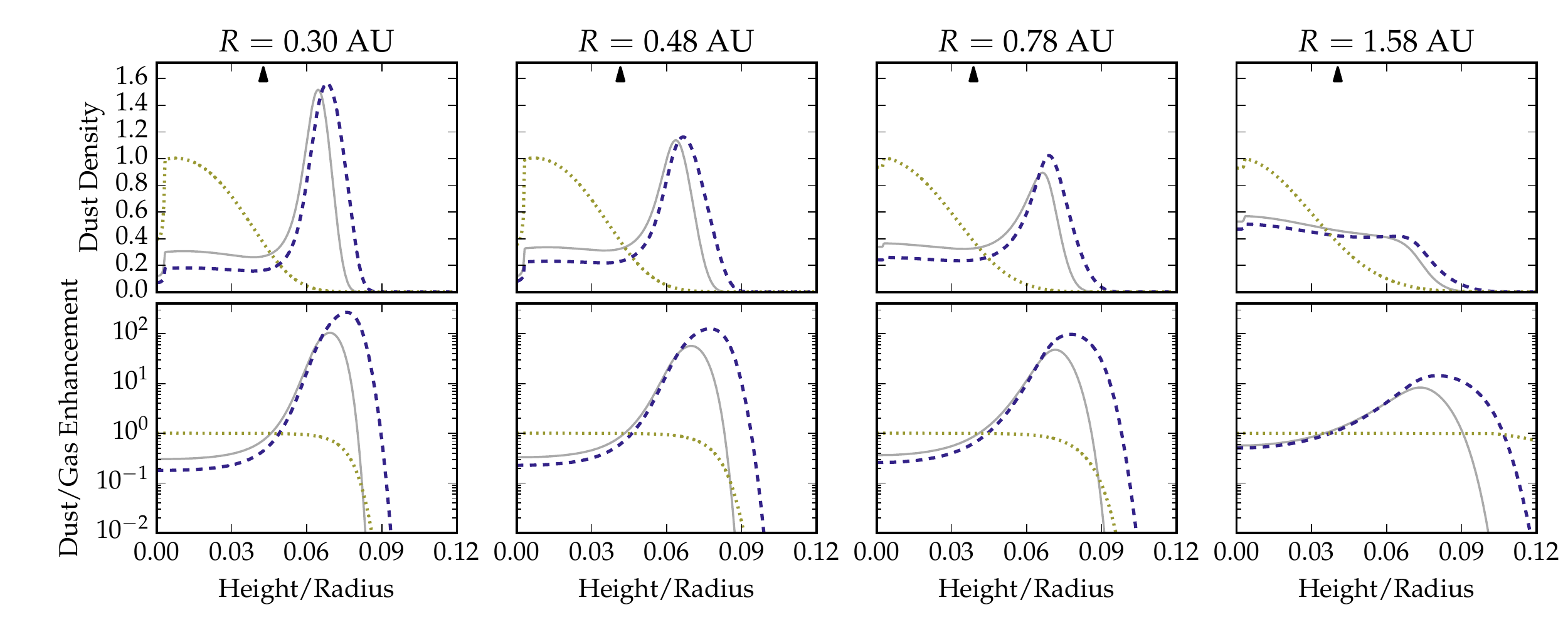}  
\caption{Particle radius variation, $a=2.5\times10^{-3}\ \mathrm{cm}$ case:  equilibrium vertical distribution of dust particles with $\phi=1$, with ${\rm Sc_0}=1.5$. 
Top row: density relative to the well-mixed peak density.
Bottom row: dust to gas density ratio enhancement factor over the well-mixed density. Green dotted line: turbulence and gravity only. Blue dashed line: including photophoresis. Gray solid line: fiducial case.}
\label{fig_a5e-4_phi1_sc1p5}
\end{figure*}

\begin{figure*}
\includegraphics[width=\textwidth]{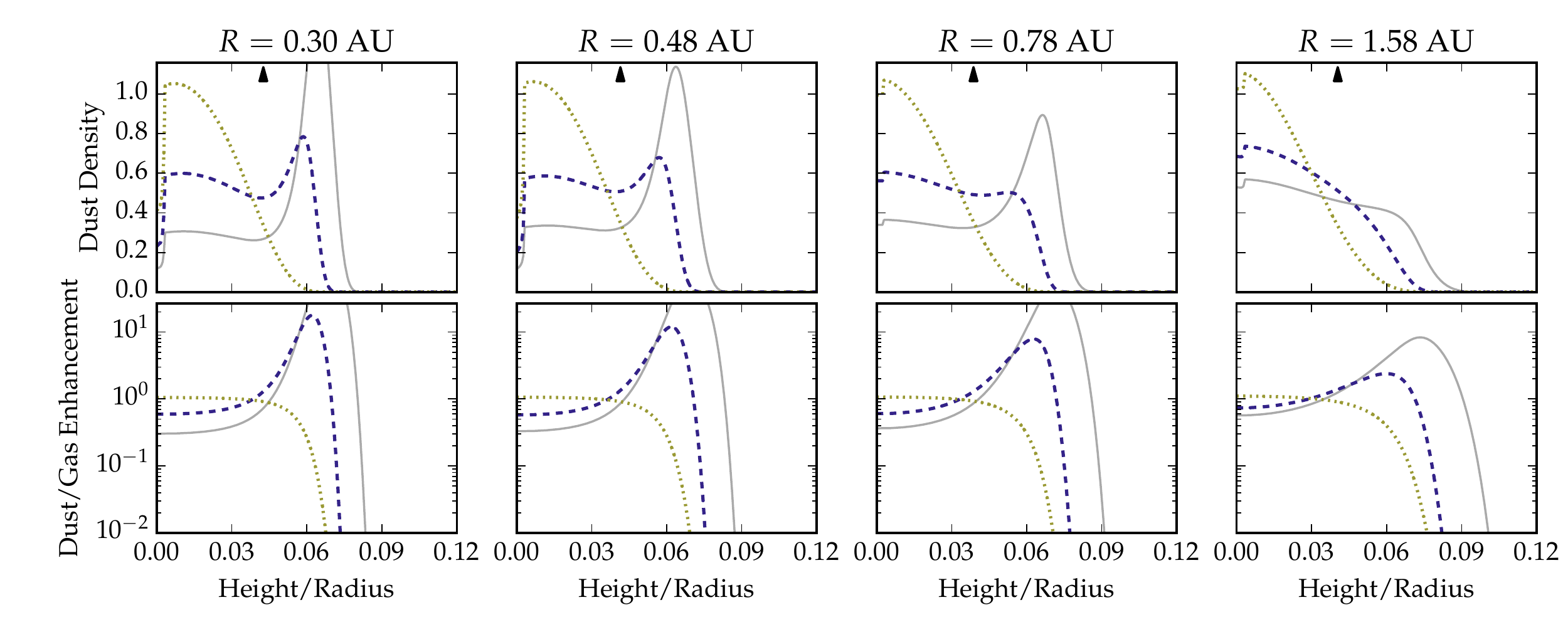}  
\caption{Particle radius variation, $a=5\times10^{-2}\ \mathrm{cm}$ case: equilibrium vertical distribution of dust particles with $\phi=1$, with ${\rm Sc_0}=1.5$. 
Top row: density relative to the well-mixed peak density, the black triangle marks one gas density scale height.
Bottom row: dust to gas density ratio enhancement factor over the well-mixed density. Green dotted line: turbulence and gravity only. Blue dashed line: including photophoresis. Gray solid line: fiducial case.}
\label{fig_a5e-2_phi1_sc1p5}
\end{figure*}

\begin{figure*}
\includegraphics[width=\textwidth]{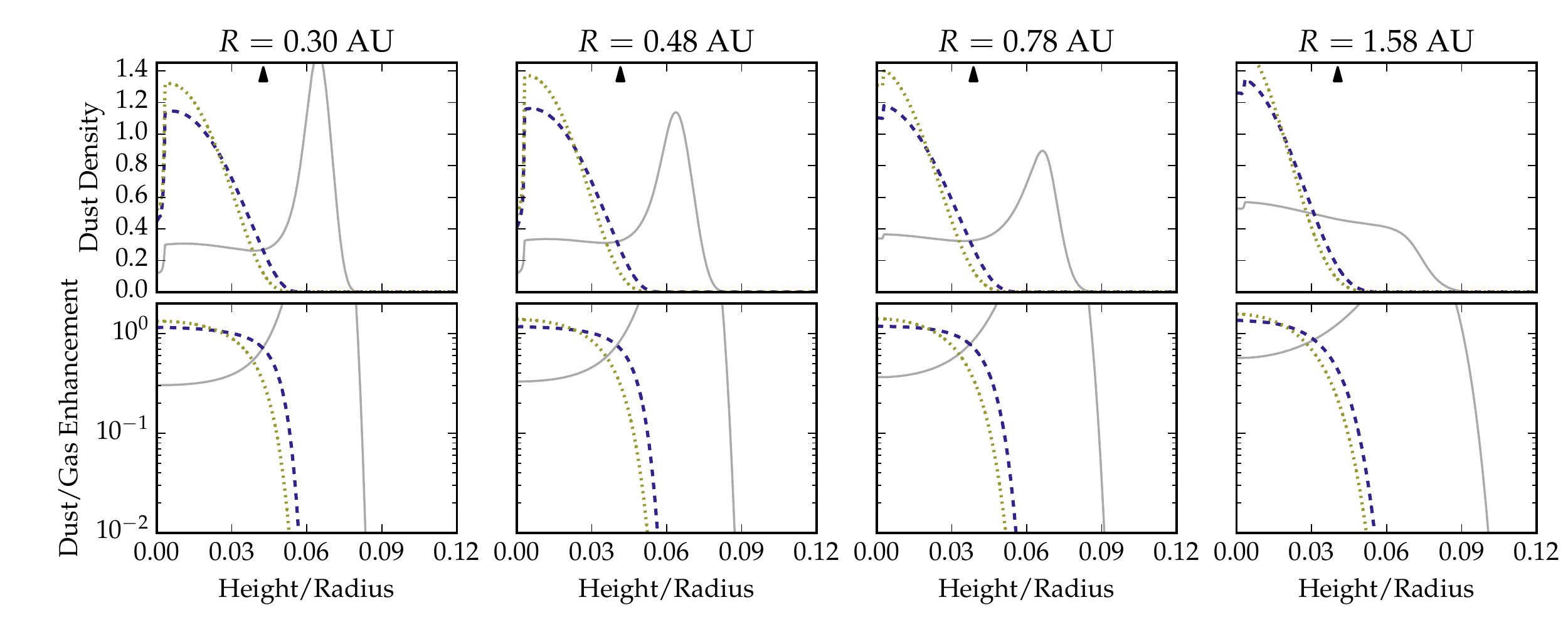}  
\caption{Equilibrium vertical distribution of dust particles with $a=5\times10^{-1}\ \mathrm{cm}$, $\phi=1$, with ${\rm Sc_0}=1.5$. Top row: density relative to the well-mixed peak density, the black triangle marks one gas density scale height.
Bottom row: dust to gas density ratio enhancement factor over the well-mixed density. Green dotted line: turbulence and gravity only. Blue dashed line: including photophoresis. Gray solid line: fiducial case.}
\label{fig_a5e-1_phi1_sc1p5}
\end{figure*}

The outcome of photophoretic levitation varies importantly with particle size.
The approximation, that incident photons deposit their energy on the side of the particle they encounter, 
begins to break down for particles with size of order the wavelength of the light.
Here, we vary particle size, but defer the detailed treatment of the regime, where 
photophoresis is gradually cut off for small particle sizes, for a future work.
As the peak wavelength for $100~\mathrm{K}$ thermal radiation is $2.9\times10^{-3}\mathrm{\ cm}$,
the smallest particles we consider have radius $2.5\times10^{-3}\ \mathrm{cm}$, a quarter the size  of the fiducial case 
shown in 
Figure~\ref{fig_a5e-4_phi1_sc1p5}.
Thus,  we expect the lower limit on the size of particles photophoretically 
levitated to be set by the characteristic wavelength of the incident radiation field, as 
opposed to the relative scaling of the diffusive and photophoretic effects with particle size.

At the other end of the grain size spectrum, although the photophoretic force increases with particle radius, 
so does the particle mass, making dust settling more effective. Figures~\ref{fig_a5e-2_phi1_sc1p5} and~\ref{fig_a5e-1_phi1_sc1p5} 
show the effect of increasing the grain sizes from the fiducial value to $a=5\times10^{-2}\ \rm{cm}$ and $5\times10^{-1}\ \rm{cm}$.
From Figure~\ref{fig_a5e-1_phi1_sc1p5}, we conclude that in the case of CI~Tau, 
the effective cutoff in particle size for the effect of photophoretic 
 levitation in the inner disk is a fraction of a centimeter.

The size dependence of dust photophoretic levitation introduces a strong size sorting effect.
In the inner disk, in addition to the enhancement in the photophoretic trap,
 the midplane dust is relatively depleted in
 particles of these intermediates sizes, due to the effect.

\subsection{Particle Porosity}

\begin{figure*}
\includegraphics[width=\textwidth]{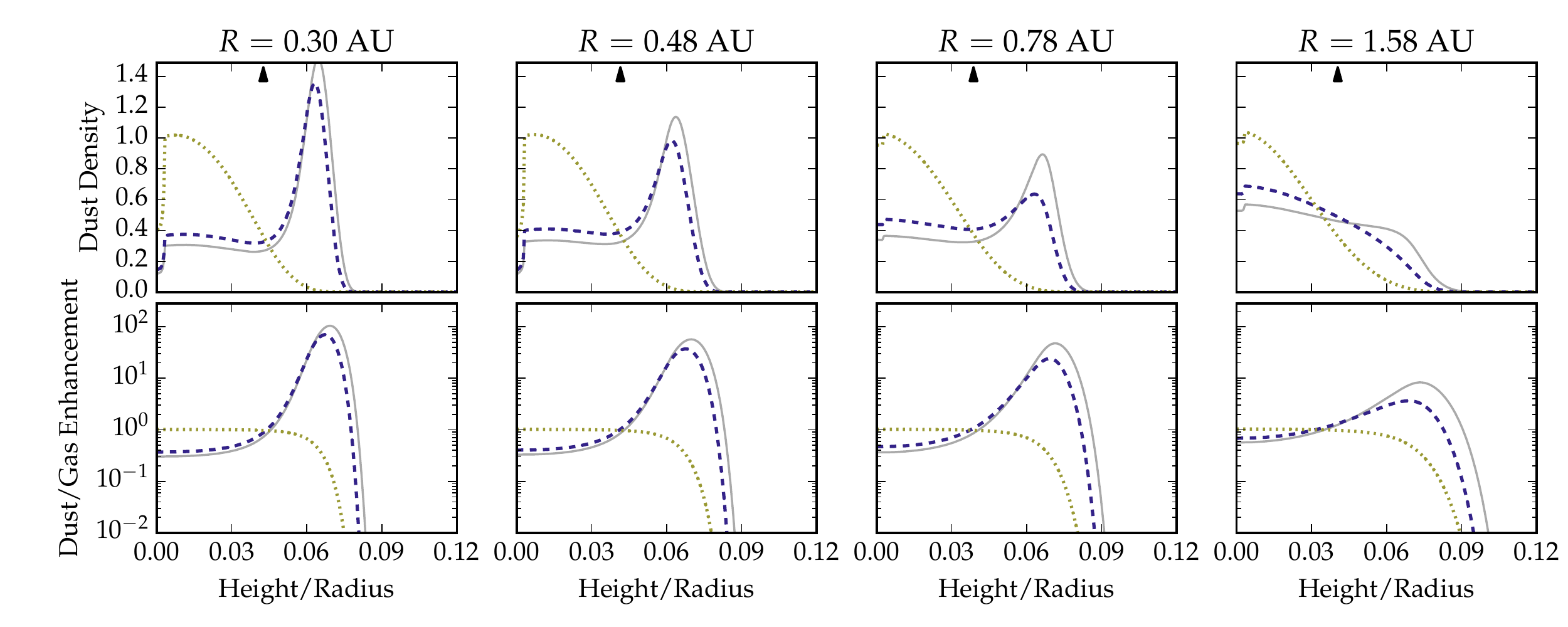}  
\caption{Particle porosity variation, $\phi=0.3$, $a=5\times10^{-2}\ \mathrm{cm}$ case: equilibrium vertical distribution of dust particles with ${\rm Sc_0}=1.5$. Top row: density relative to the well-mixed peak density, the black triangle marks one gas density scale height.
Bottom row: dust to gas density ratio enhancement factor over the well-mixed density. Green dotted line: turbulence and gravity only. Blue dashed line: including photophoresis. Gray solid line: fiducial case.}
\label{fig_a5e-2_phi0p3_sc1p5}
\end{figure*}

\begin{figure*}
\includegraphics[width=\textwidth]{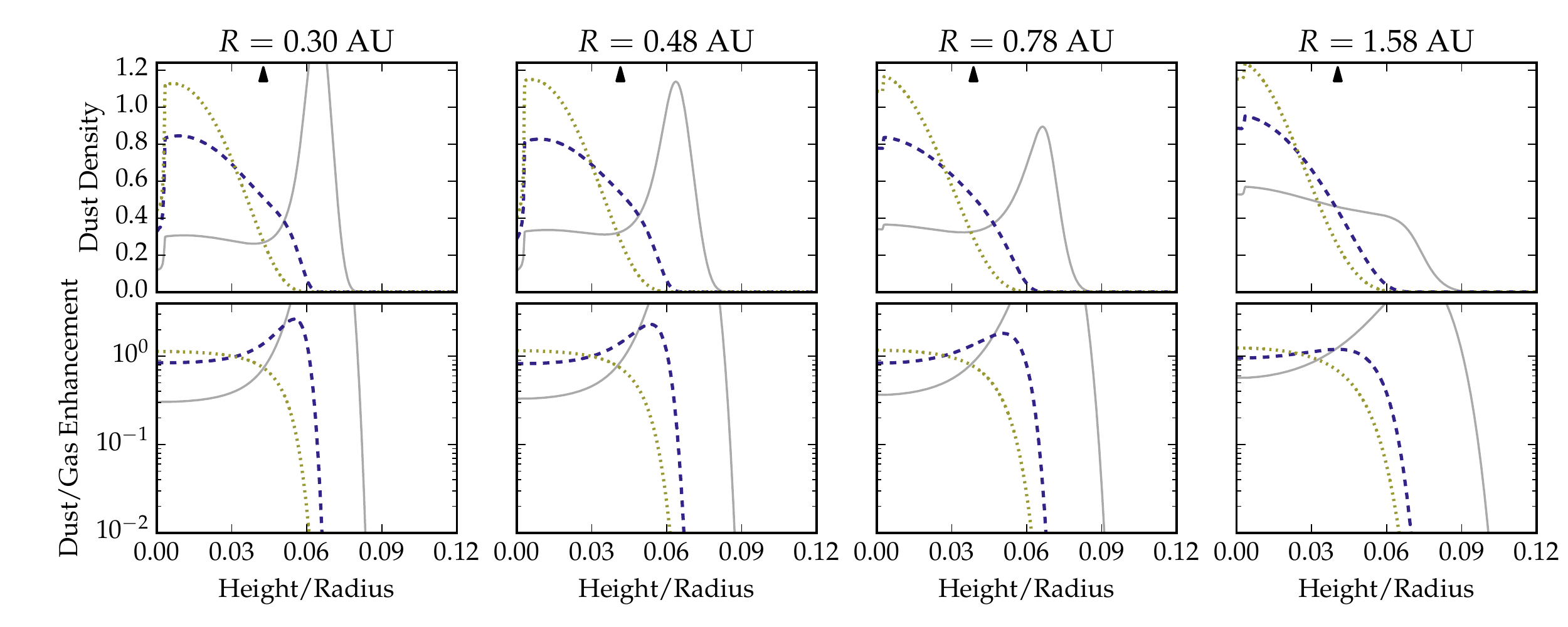}  
\caption{Particle porosity variation, $\phi=0.3$, $a=5\times10^{-1}\ \mathrm{cm}$ case:  equilibrium vertical distribution of dust particles with ${\rm Sc_0}=1.5$. Top row: density relative to the well-mixed peak density, the black triangle marks one gas density scale height.
Bottom row: dust to gas density ratio enhancement factor over the well-mixed density. Green dotted line: turbulence and gravity only. Blue dashed line: including photophoresis. Gray solid line: fiducial case.}
\label{fig_a5e-1_phi0p3_sc1p5}
\end{figure*}

\begin{figure}
\includegraphics[width=\columnwidth]{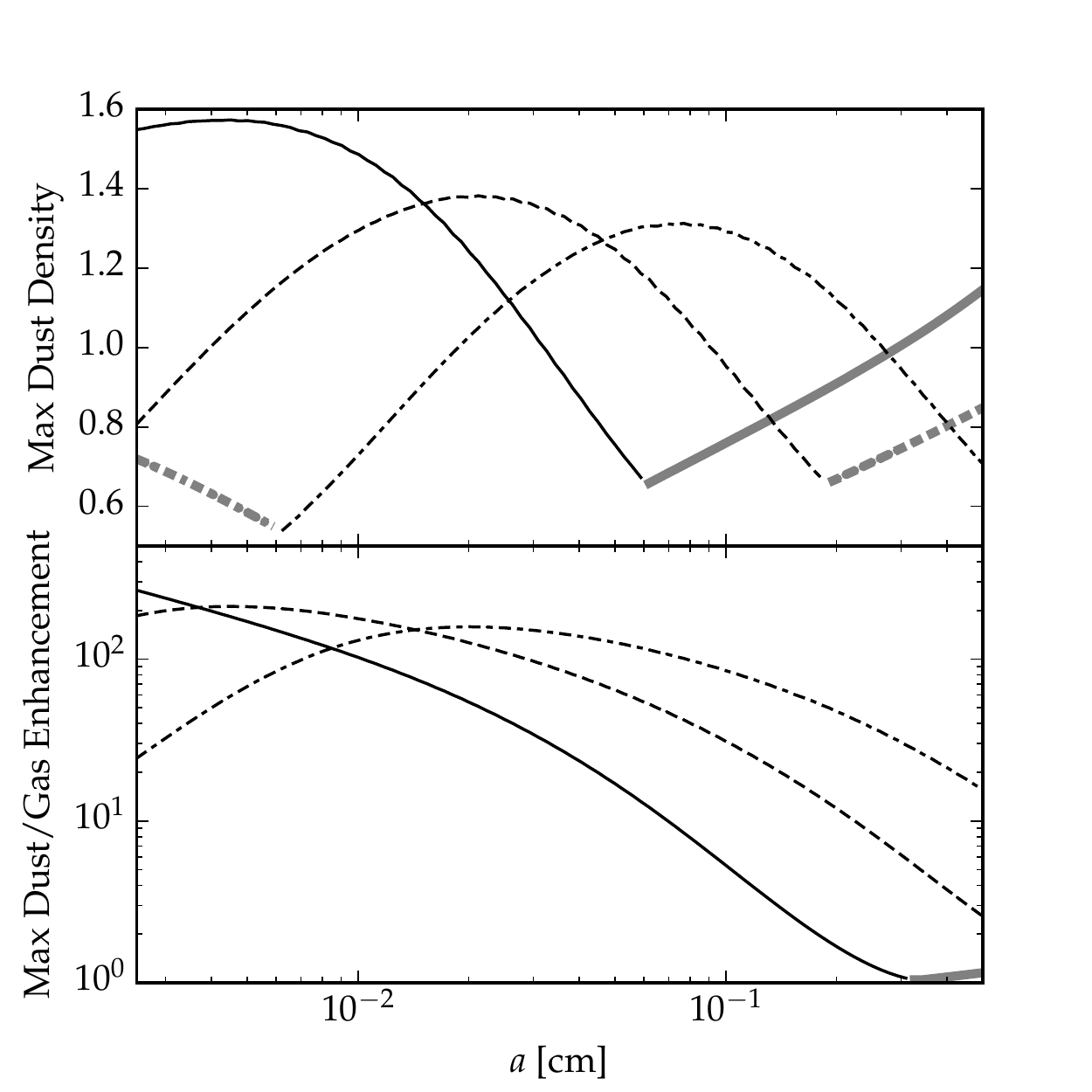}
\caption{ 
CI~Tau parameter scan at $R=0.3 \ \mathrm{au}$ for varying particle 
radius $a$ at three porosity values $\phi$. 
Top: maximum dust density.  
Bottom: dust/gas enhancement.
Solid line: $\phi=1$, Dashed line: $\phi=0.3$, Dash--dot line: $\phi=0.1$,
Thick gray line: the maximum of the respective value is located at the midplane, not in the photophoretic trap layer.}
\label{fig_citauscan}
\end{figure}

It is likely that silicate grains in the inner disk have a non-trivial porosity.
The variation of particle porosity changes both the density of the grains and the thermal conductivity.
In this section, we examine the effect of varying the particle porosity following the aggregate model
used in \citetalias{2015ApJ...814...37M}.

The most interesting regime is for large particles with significant porosity,
so in Figure~\ref{fig_a5e-2_phi0p3_sc1p5} the result for particles
with radius $5\times10^{-2}\ \rm{cm}$ and volume filling factor $\phi=0.3$ is shown, which can be compared to
the result for solid ($\phi=1$) particles in shown in Figure~\ref{fig_a5e-2_phi1_sc1p5}.
The particle trapping effect is drastically increased for those particles.
For yet larger particles, with  radius $a=5\times10^{-1}\ \rm{cm}$ 
 and $\phi=0.3$ shown in Figure~\ref{fig_a5e-1_phi0p3_sc1p5} 
 (compare with Figure~\ref{fig_a5e-1_phi1_sc1p5})
 the effect of increasing porosity is less dramatic, although still present,
 as it introduces a local peak in the dust/gas enhancement in the trap.
 Lofting porous particles to several scale heights in this disk model presents a 
 challenge as the collisional velocity at height can easily be 
above the fragmentation velocity proposed by \citet{2011A&A...525A..11B}.

We note that the results depend in some regimes on particle porosity, but varying the density of the 
underlying silicate monomers in the reasonable range of $2.5-3.5\ \mathrm{g\ cm^{-3}}$
does not drastically affect the results considered here.

\subsection{Combination of Size and Porosity}
In this section, we survey a larger number of parameter values, 
but restrict our attention to a single radial position at $R=0.3\ \mathrm{au}$.
Plotting the maximum of the dust density over than range in 
Figure~\ref{fig_citauscan}, in the upper panel, shows two important features.
First, the local minima correspond 
to the minimum dust density in the photophoretic trap, falling below the increasing 
dust density at the midplane, as the photophoretic 
lofting becomes less effective.
The left hand end of the curves, in particular the $\phi=0.3$ and $\phi=0.1$ cases, shows
the photophoretic trapping being limited by the diffusion of particles out 
of the trap - as the stopping time of the small fluffy particles decreases, the 
maximum dust density in the trap decreases as turbulence diffuses them out.
In the corresponding plot of the maximum dust/gas enhancement 
(Figure~\ref{fig_citauscan}, lower panel)
a local minimum is only seen in the $\phi=1$ curve, and here again it corresponds to 
the location of the maximum value transitioning from the dust trap to the midplane.
In general, these figures show that the effect of photophoretic trapping is 
shifted to geometrically larger particles as the porosity increases.

\subsection{Levitation/Settling Timescale}

\begin{figure}
\includegraphics[width=\columnwidth]{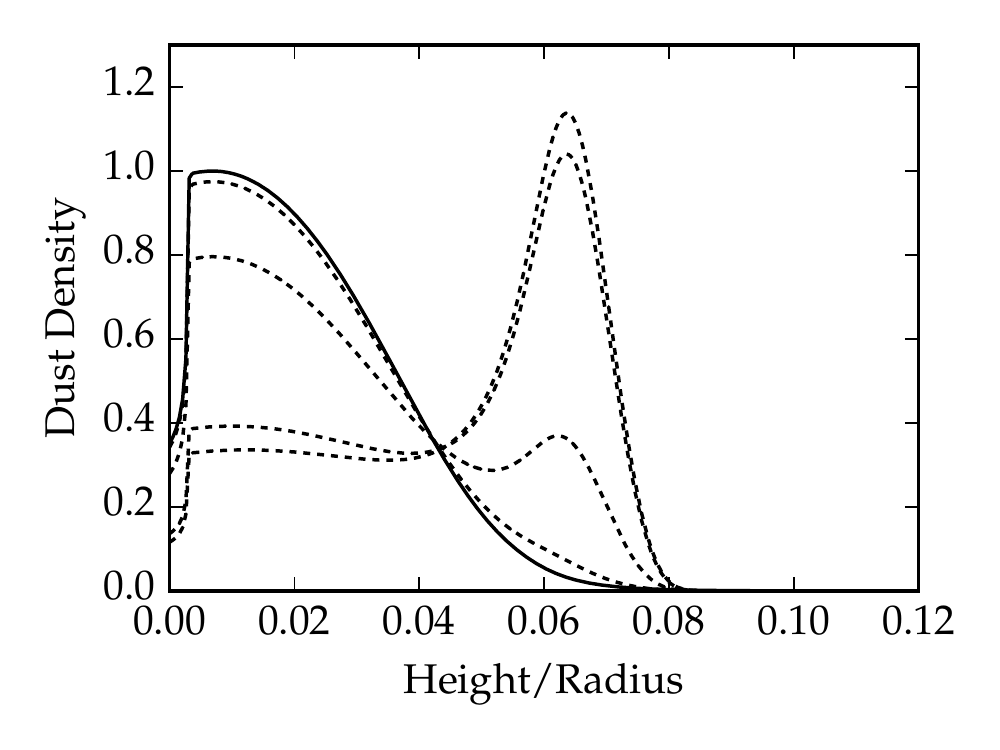}
\caption{Time evolution of the distribution of dust particles with $a=1\times10^{-2}\ \mathrm{cm}$, 
$\phi=1.0$, with ${\rm Sc_0}=1.5$ at $R=0.48\ \mathrm{au}$. 
Solid curve: Initial well-mixed distribution. 
Dashed curves: in order from the solid curve, times $1$, $10$, $10^2$, $10^3$ years.}
\label{fig_timescale}
\end{figure}

In Figure~\ref{fig_timescale} the time evolution of the $R=0.48\ \mathrm{au}$ 
column of the fiducial case is given, showing that the equilibirum dust distribution 
is established on a timescale on the order of $100$ years.
As in the case of purely gravitational settling \citep{2004A&A...421.1075D}, the vertical equilibrium is
established on a timescale much shorter than the accretion timescale.

\section{The Case of V836 Tau}
\label{sec_v836}

\begin{figure}
\includegraphics[width=\columnwidth]{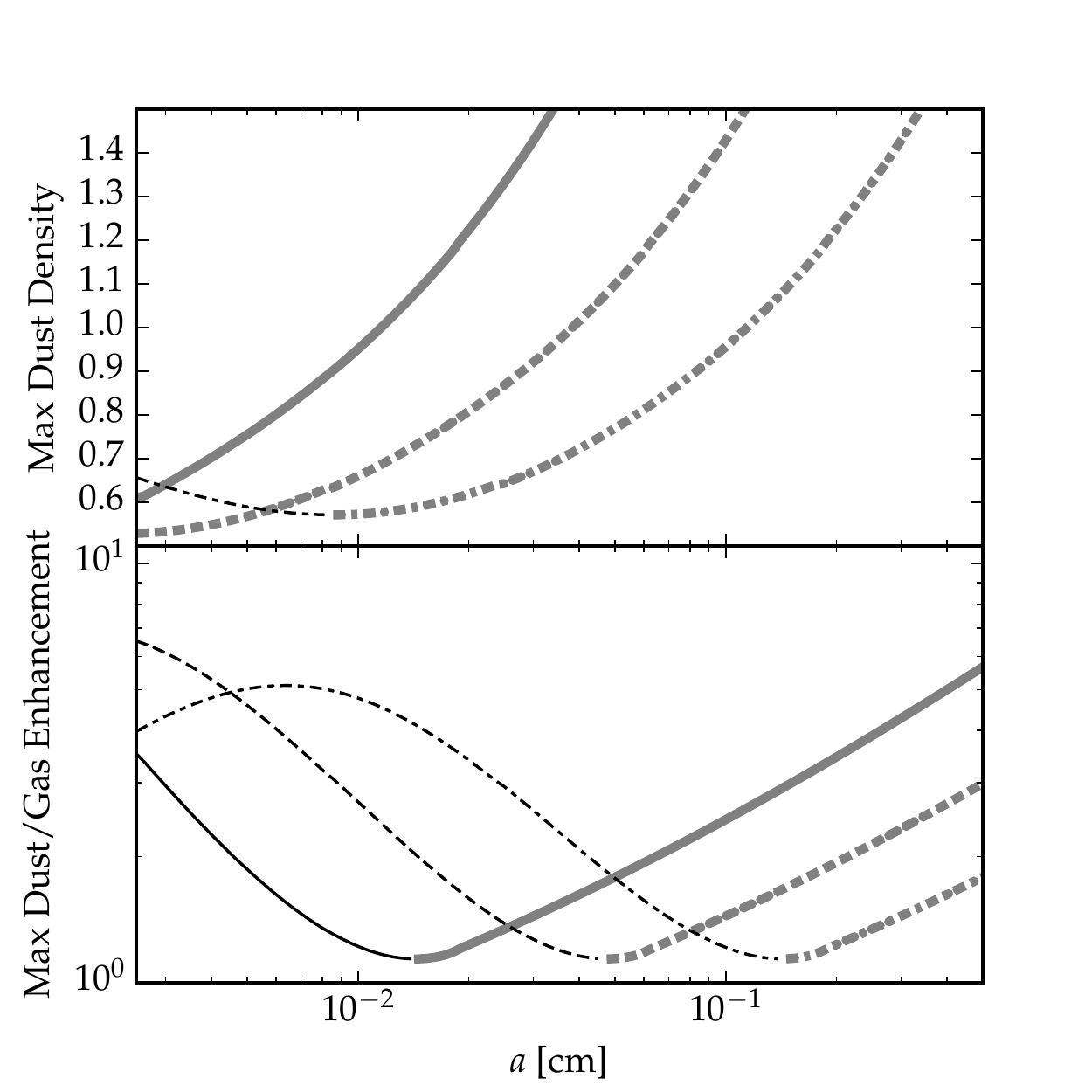}
\caption{ 
V836 Tau parameter scan at $R=0.25 \ \mathrm{au}$ for varying particle 
radius $a$ at three porosity values $\phi$. 
Top: maximum dust density.  
Bottom: dust/gas enhancement.
Solid line: $\phi=1$, Dashed line: $\phi=0.3$, Dash--dot line: $\phi=0.1$,
Thick gray lines: the maximum of the respective value is located at the midplane, not in the photophoretic trap layer.}
\label{fig_v836scan}
\end{figure}

As a comparison to CI~Tau, the V836~Tau disk has a much less effective photophoretic levitation effect.
V386 Tau is again a classical T~Tauri system, with the important characteristics of a much lower accretion rate
 ($\dot{M}= 1.9\times10^{-1}\ {\rm M_\sun\ yr^{-1} }$) and a much lower implied turbulent 
 viscosity ($\alpha_{\rm ss} = 8\times10^{-5}$, \citetalias{2013ApJ...775..114M}).
 Thus the disk has a lower surface density and less effective accretion heating in the inner disk.
The maximum dust denstiy and dust/gas enhancement in a column at
 at $R=0.25\ \mathrm{au}$ where the midplane temperature in the 
 model is $420\ \mathrm{K}$ is shown in Figure~\ref{fig_v836scan}, and can be 
 compared to a similar column in CI Tau shown in Figure~\ref{fig_citauscan}.
 In the maximum dust density panel, the curves rise monotonically to the right, 
 as a result of the maximum dust density always occurring at the midplane, 
 and not  in the photophoretic trap.
In the lower panel showing the maximum dust/gas enhancement, the sections of the curves to the 
left of the minima indicate the maximum enhancement occurring in the photophoretic trap layer.
In the case of V836, the photophoretic trapping effect can only be said to be mild, 
and even then so for the smallest and most porous grains.
This indicates the role of the specific disk properties in determining the efficacy of photophoretic levitation.

\section{Discussion}
\label{sec_disc}

We have found in one-dimensional dust settling calculations that in this disk model the dust density 
maximum occurs far above the disk midplane when the photophoretic force on the dust grains is included, 
in contrast to considering gravity and dust turbulent diffusion alone.
Dust is lifted away from the midplane by photophoresis and stops rising 
when the vertical components of gravity become larger than the photophoretic force,
due to the rapidly decreasing gas density of the disk atmosphere.
This effect occurs only in the inner regions of the disk, where the 
accretion heating in the midplane and the vertical flux of energy released is significant.

The upper height of the photophoretic layer is effectively set by the decreasing gas density,
or in other words the increasing Stokes number of the lofted particles.
Because of this, the large particles are not found to be lofted above 
the very small dust grains providing the bulk of the disk opacity.
Thus, although grains are lofted, the configuration proposed by
\citet{2009M&PS...44..689W}
where photophoretically levitated grains enter the fully optically thin
 atmosphere above an FU~Orionis-state disk and are directly exposed to the stellar 
 irradiation, is not realized. This suggests that further work on that model 
 should consider the settling and opacity caused by micron-sized grains.

\subsection{Directions for Further Development}
\label{sec_further}
Several details concerning the photophoresis of real dust particles have not been treated in this work.
Importantly, the treatment of photophoresis used in this work considers only grains 
of an idealized type, while the underlying theory of photophoresis should 
be extended to include
dust grain sizes comparable to the wavelength of the disk's thermal emission,
significantly nonspherical grains,
and  grains with very large porosity and/or non-silicate compositions.

As discussed in Section~\ref{sec_preconsistency} the models in this work can 
only be considered to be fully self-consistent if
the concentration of the large grains which feel photophoresis 
does not significantly affect the disk's opacity.
At very high concentrations this may not be justified. 
Indeed, when the fragmentation of large grains in the trap layer is
 considered there is a particular possibility for the local opacity to be modified.
In this case, it would be nescessary to develop a full model for the population 
of dust grains of each kind, by size, composition, and porosity; solve a trial vertical settling 
problem for each dust species; 
iterate with this dust distribution for the radiative-hydrostatic balance within the disk column; 
and then iterate, re-solving the vertical settling problem and radiative-hyrostatic balance problem until converged.
To our knowledge, such a calculation has not been performed in the literature, even neglecting photophoresis and
 dust fragmentation-coagulation driven evolution.
Similarly, the  turbulent diffusion approximation for the dust motion
is expected to break down  when the dust to gas mass ratio is locally high enough \citep{1995Icar..114..237D,2016arXiv160302630K}.
This may be achieved in the photophoretic dust trap layer if the 
absolute abundance of trapped grains is sufficiently high.

Localized temperature fluctuations in the disk gas
can be produced by magnetohydroynamic turbulence
 \citep{2011ApJ...732L..30H,2014ApJ...791...62M}.
Thermal radiation from these local hot spots locally may produce 
photophoresis in turn \citep{2015LPICo1856.5137L,2015ApJ...814...37M}.
As these fluctuations are localized and randomized, they are expected to be similar in 
nature to the effect of turbulence on the diffusion of the dust, and we consider 
the effect on the results here to be subdominant to the uncertainty in the Schmidt number.

The disk model used here is one calibrated to match observations, but dynamical models of protoplanetary disks
contain features and restrictions beyond what is included.
Of greatest potential significance are the possibilities of wind driven accretion and the 
details of the turbulence which is modeled as a viscous $\alpha$.
If the observed accretion rate of the system is driven by a
 disk wind torque \citep{1983ApJ...274..677P,2015ApJ...801...84G,2016ApJ...821...80B}, 
then the inferred value of 
$\alpha$ used in this paper may overestimate the turbulent energy dissipation, 
and hence thermal radiation flux driving levitation.
The disk model used here assumes a constant value of the turbulent $\alpha$,
whereas the relevant properties real turbulence in the disk may not have the same 
simple spatial dependence. 
Both the variation with radius 
and height may be different, for example due to thermodynamic or non-ideal MHD effects
\citep{2009ApJ...701..737B,2011ApJ...732L..30H}.
Thus, the complete understanding of dust motion driven by photophoresis in these systems ultimately demands 
 a more comprehensive inclusion of the physics of protoplanetary disks than is currently available.
At the same time, as photophoretic levitation is sensitive to these details of the interior workings of the disk,
in particular the critical question of whether the accretion at small radii is driven by winds or turbulent stresses,
the presence or absence of the effect may provide a way of testing disk models.

\subsection{Effects on fragmentation and grain growth}
The redistribution of large grains has significant implications for fragmentation and grain growth. 
Grain collisions driven by turbulence limit the grain size in a very porosity-dependent way, 
as solid silicate grains are much tougher than porous aggregates.
In the $\alpha$-turbulence model used in this work, the grain-grain collision velocity,
if dominated by the turbulent motions, is of the order of $\sqrt{\alpha_{\rm ss}\, {\rm St}}\, c_s$,
and as the Stokes number of particles increases with the decreasing density of the disk 
atmosphere, and the sound speed increases with temperature, the collisional velocities are
 expected to grow with height \citep{2009A&A...503L...5B}.
Hence, for some porous grains the levitation could result in an increased fragmentation rate. 
However, the mass ratio between colliding particles will be enhanced in both the midplane and
 mid-altitude regions by shifting the distribution of $\sim10^{-2}$ cm sized grains. With fewer collisions between 
 grains of similar sizes, there should be less fragmentation even at relatively high velocities \citep{2012A&A...540A..73W}. 
 This may lead to improved 
 grain growth relative to a settling-only case. If even a few large grains survive, they 
 can break through the bouncing and fragmentation barriers to become seed 
 particles for planetesimal formation \citep{2012A&A...544L..16W}.
The scenario for the local concentration of the small dust population associated with the 
fragmentation of large grains at the midplane proposed by \citet{2016arXiv160302630K} would also
be altered in the presence of a photophoretic trap or even lofting effect altering the vertical distribution 
and collision rates of the large grains.

As in this work, we have generalized vertical settling of a single particle species to include photophoresis,
the natural extension is to generalize the coupled settling-coagulation-fragmentation 
problem in the same way \citep{2005A&A...434..971D}.
Changes to the production and distribution of small grains have the potential to alter the opacity of the disk,
and have potential effects on the SED.

\subsection{Impact of trap on disk dynamics}

The vertical dust trap indicated by the one-dimensional calculations in this work likely has
important consequences when considered in higher dimensions and in fully dynamic models.
As the concentration can be on the order of one hundred times well-mixed for large particles,
the relative dust/gas enhancement can approach $10^2$ times the global dust/gas mass ratio.
This enhancement suggests local dust/gas mass ratios of at least unity in the trap, 
or more if dust filtration in the disk has produced 
a higher vertically averaged dust/gas mass ratio at these inner radii. 
At this level, the dust momentum is no longer
negligible with respect to the gas momentum, and instabilities are likely.
Although as dust is lofted, the Stokes number of the particles increases due to the decreasing density.
The Stokes number reached by particles described in this work is well below ${\rm St}=0.02$,
and typically in the range $10^{-3}$--$10^{-4}$
 for the vast majority of the dust mass in the photophoretic dust trap.
 Thus, by the results of \citet{2015A&A...579A..43C} one would not expect the layer
 to break up due to streaming instability.
 If streaming instability can be demonstrated 
 to develop at Stokes numbers in the order of $10^{-3}$, 
 then the photophoretic dust trap layer might provide a way of rapidly transforming 
several-hundred-micrometer dust directly into much larger aggregates,
 bypassing barriers in the coagulation growth of aggregates in the inner disk.
We intend to further explore the stability of the dust trapping layer in future work.

These models are viscous disks, which at the temperatures considered 
are in tension with the predictions for dead zone formation from 
magnetohydrodynamic models \citep{2013ApJ...765..114D}.
They are, however, agnostic about the underlying physical 
mechanism giving rise to the $\alpha$-turbulence for dust 
stirring and the $\alpha$-viscosity leading to energy dissipation.
An extension of this work to similar 1+1D hydrostatic disk models,
which take into account the parameter dependence of dead zones,
is worthwhile \citep{2013ApJ...771...80L,2016arXiv160404601F}.
If the photophoretic lofting heat source due to 
turbulence were to switch off due to a change in the
nature of the turbulent accretion (such as the end of a
gravitational instability driven dead-zone bursting period),
the settling of the dust layer may drive clumping
through the instability proposed by \citet{2016arXiv160400791L}.

\subsection{Implications for the formation of chondrites and planets}
The range of sizes for solid silicate particles which are 
strongly trapped is very interesting in the context of meteoritical studies.
In a review of analyses of many meteorites, \citet{2015ChEG...75..419F} 
show the distribution of chondrule radii in  H, L, and LL chondrites 
typically peaks near $250\ \mathrm{\mu m}$.
Thus, these chondrules (reasonably approximated as solid silicate spheres)
would be strongly levitated and trapped in a disk like CI~Tau.

Calcium-aluminum-rich inclusions, if produced early in 
the inner part of the solar nebula, would be processed by 
temperature fluctuations occurring during their turbulent transport
 \citep{2013LPI....44.2007T,  2015Icar..252..440C}.
The particle sizes, disk epoch and radii suggest 
photophoretic levitation has a significant effect on these 
histories as it will drive them away from the midplane.

Likewise, the formation of  STIPs (Systems of Tightly Packed Inner Planets)
 has been proposed to rely on the migration and pileup of dust at small radii
 \citep{2012ApJ...751..158H,2014ApJ...792L..27B}.
Indeed, even CI~Tau shows radial velocity variations consistent with a hosting a
 hot Jupiter class planet \citep{2016arXiv160507917J}.
These  radii overlap with those where we predict the formation of the photophoretic dust trap.
Trapping grains in this layer results in a lowering the density of the same grains at the 
midplane, and the stability of the trapped dust layer should be explored for the 
potential consequences to planet formation at small radii.

\section{Conclusions}
\label{sec_conc}
In this work we present a treatment of photophoresis using a two-stream radiative transfer approximation and 
demonstrate that the thermal radiation of the disk itself can levitate large dust grains. By combining photophoresis, turbulence, and 
dust settling calculations, we obtain a clearer view of the vertical distribution of large dust grains in the inner 2~$\mathrm{AU}$ of protoplanetary disks.
The application of this approach to an observationally calibrated disk structure model that includes heating by both viscous dissipation 
and stellar irradiation processes, for ``typical'' T~Tauri star disks, yields the following conclusions:
\begin{enumerate}
\item  The addition of the photophoretic force to the dust settling calculation creates a strong vertical dust trap above the 
disk midplane, at altitudes under the stellar irradiation surface.
\item The peak dust/gas enhancement in the trap approaches $10^2$ times the vertically averaged dust/gas mass ratio
for the respective particle type. 
\item The degree of enhancement depends strongly on the grain size considered and the particle porosity.
\end{enumerate}

The presence of a significant enhancement in the dust/gas ratio above the midplane, as well as the redistribution of large grains, may 
produce interesting dynamical effects and will affect grain growth and fragmentation patterns in the innermost disk. 
Understanding the limits of this mechanism and its implications for planet formation in this region 
will be the focus of future work.

\acknowledgements
We acknowledge useful discussions with Alexander Hubbard, 
Zhaohuan Zhu, James Owen, Chao-Chin Yang, Henning Haack, and Christian Brinch.
The research leading to these results has received funding from the 
European Union's Seventh Framework Programme 
(FP7/2007-2013) under  ERC grant agreement 306614 (CPM).

\appendix
\label{sec_spin}
\section{Dust Particle Spin}

In the interior of the protoplanetary disk, a spherical dust grain in thermal equilibrium with the surrounding gas
has a rotation rate set by the $3/2 k_B T_g$ energy associated with the rotational degree of freedom of \citep{2008ipid.book.....K}
\begin{align}
\omega_\mathrm{brownian} = \sqrt{\frac{45 k_B T_g}{8 \rho_d}} a^{-5/2} \ . \label{eq_thermal_spin}
\end{align}
This rotation rate has a steep dependence on the partical radius $a$. 
For a temperature at the high  end of the applicable range, $1200\ \mathrm{K}$,
the rotation period of a grain with density $\rho_d = 3 \ \mathrm{g\ cm^{-3}}$ and radius $a=2.5\times 10^{-3}\ \mathrm{cm}$ 
is about $3.5\ \mathrm{s}$, whereas for a $a=1\times 10^{-1}\ \mathrm{cm}$ grain it is
about $3.5\times10^4\ \mathrm{s}$. 
Under these conditions, the angular velocity kick to the smaller grain considered due to a collision with
 a single ${\rm H}_2$ molecule is of the order of $10^{-8}\ \mathrm{rad\ s^{-1}}$ and the disorder 
 time of the brownian motion is on the order of $20\ \mathrm{s}$ \citep{2008ipid.book.....K}.

For the thermal conduction time across the grain, we can follow the estimate 
provided by \citet{2016MNRAS.455.2582M} of
\begin{align}
\tau_\mathrm{heat} = \frac{\rho_d c_d a^2}{k}\ .
\end{align}
Taking the  heat capacity $c_d= 10^7 \ \mathrm{erg\ g^{-1}\ K^{-1}}$,
thermal conductivity $k=1.4\times 10^{5} \ \mathrm{erg\ s^{-1} \ cm^{-1} \ K^{-1}}$
yields conduction times of $1.3\times 10^{-3}\ \mathrm{s}$ for the $a=2.5\times 10^{-3}\ \mathrm{cm}$  grain and
$2.1\ \mathrm{s}$ for a $a=1\times 10^{-1}\ \mathrm{cm}$ grain.
The closest pair of rotation period and conduction times for particles in the 
parameter space considered here at $1200\ \mathrm{K}$, is 
for a grain with filling factor $\phi=0.11$ and $a=2.5\times 10^{-3}\ \mathrm{cm}$
where the rotation period is $1.1\ \mathrm{s}$ and the conduction time is $0.16\ \mathrm{s}$.
Hence, in general, we expect the temperature gradient that drives photophoresis to be 
sustained against the rotation of dust particles as the heat conduction times are short 
compared to the other timescales in the problem. 

In addition to collisions with gas molecules which bring the dust spin into thermal equilibrium with the gas,
dust--dust collisions can spin up dust to much higher rates.
Here, we determine an order-of-magnitude criteria for the 
spin rate in equation~(\ref{eq_thermal_spin}) to dominate the time-averaged spin of a dust particle.
Assume that the dust collision velocities are in the order of the turbulent velocities $\sqrt{\alpha_{\rm ss}\, {\rm St}}\, c_s$
 \citep{2009A&A...503L...5B}, then the angular momentum imparted to a 
 dust grain from the worst-case equal size collision would be
 in the order of
\begin{align}
\Delta L_d \sim m_d a \sqrt{\alpha_{\rm ss}\, {\rm St}}\, c_s\ .
\end{align}
For a spherical grain the surface rotational velocity kick after the collision would be
\begin{align}
\Delta v \sim \frac{5}{2} \sqrt{\alpha_{\rm ss}\, {\rm St}}\, c_s\ .
\end{align}
The angular momentum which is removed from this spinning dust particle to the thermal bath provided by the gas 
by a subsequent collision with a gas molecule is on the order of
\begin{align}
\Delta L_g \sim m_g a \Delta v = m_g a \frac{5}{2} \sqrt{\alpha_{\rm ss}\, {\rm St}}\, c_s\ .
\end{align}
So the spinning dust will have slowed to a thermal equilibrium after  a number of the order of 
$N$ collisions where
\begin{align}
\Delta L_d &= N \Delta L_g \ ,\\
m_d &= \frac{5}{2} m_g N\ .
\end{align}
So, as long as the total mass of gas colliding with the spinning dust particle per unit time is much larger than the total 
mass of other dust particles colliding with it, then the spin will not wander far from equilibrium in a time-averaged sense.
We define this ratio of colliding gas to dust as $\Gamma$, and it can be expressed as 
\begin{align}
\Gamma \approx \frac{4 \pi v_{\rm rms} a^2}{4 \pi \sqrt{\alpha_{\rm ss}\, {\rm St}}\, c_s a^2} \chi^{-1}
\end{align}
where $v_{\rm rms} = \sqrt{3/\gamma} c_s$ and $\chi$ 
is the dust-to-gas mass ratio locally in the disk (canonically on the order of $1/100$). 
So this quantity, given that $\alpha_{\rm SS}$, $\rm St$ and $\chi$ are all small quantities, is 
\begin{align}
\Gamma \sim \frac{1}{ \sqrt{\alpha_{\rm ss}\, {\rm St}} }  \frac{1}{\chi} \gg 1 \ .
\end{align}
Hence, small particles can be expected to, in a time-averaged sense, 
typically spin at the gas thermal equilibrium rate given by equation~(\ref{eq_thermal_spin}), because they 
interact with a much larger mass of gas than of dust in a given time.

\bibliography{photolev}

\end{document}